\documentclass[10pt,conference]{IEEEtran}
\usepackage{graphicx}
\usepackage{authblk}
\usepackage[normalem]{ulem}
\usepackage{amsmath}
\usepackage[colorlinks=true,urlcolor=blue, citecolor=blue, linkcolor=blue, plainpages=false]{hyperref}   
\usepackage{amsmath,amssymb,physics,dsfont}
\usepackage{tikz}
\usetikzlibrary{quantikz2}
\usepackage{private}
\usepackage{comment}
\usepackage{mathtools, cuted}

\usepackage{adjustbox}
\usepackage{tikz}
\usepackage{quantikz} 

\usetikzlibrary{decorations.pathreplacing} 

\title{Hamiltonian Simulation for Advection-Diffusion Equation with arbitrary transport field}

\author[1]{Niladri Gomes}
\author[2]{Gautam Sharma}
\author[1]{Jay Pathak}

\affil[1]{Ansys, Inc, San Jose, CA}
\affil[2]{Ansys Software Pvt. Ltd., Pune, India}

\begin{document}
\maketitle

\begin{abstract}

We present a novel approach to solve the advection-diffusion equation under arbitrary transporting fields using a quantum-inspired ‘Schrödingerisation’ technique for Hamiltonian simulation. Although numerous methods exist for solving partial differential equations (PDEs), Hamiltonian simulation remains a relatively underexplored yet promising direction—particularly in the context of long-term, fault-tolerant quantum computing. Building on this potential, our quantum algorithm is designed to accommodate non-trivial, spatially varying transport fields and is applicable to both 2D and 3D advection-diffusion problems. To ensure numerical stability and accuracy, the algorithm combines an upwinding discretization scheme for the advective component and the central differencing for diffusion, adapted for quantum implementation through a tailored mix of approximation and optimization techniques. We demonstrate the algorithm’s effectiveness on benchmark scenarios involving coupled rotational, shear, and diffusive transport in two and three dimensions. Additionally, we implement the 2D advection-diffusion equation using 16 qubits on IBM Quantum hardware, validating our method and highlighting its practical applicability and robustness.

\end{abstract}


\section{Introduction}

Achieving quantum speedup for the numerical solution of ordinary and partial differential equations (ODEs/PDEs), beyond those specific to the Schrödinger equation, holds considerable promise for advancing a wide range of scientific and engineering disciplines \cite{yepez1998quantum,  bharadwaj2020quantum, horowitz2019quantum}. PDEs play a central role in modeling the dynamic behavior of physical systems and arise in numerous applications, including reservoir simulation, climate and ocean circulation, and wave propagation. In particular, computational fluid dynamics (CFD) relies fundamentally on the numerical solutions of PDEs that describe fluid flow phenomena \cite{wendt2008computational}. While high-performance computing (HPC) has enabled significant strides in solving such complex models, the computational cost remains a persistent challenge, especially in problems involving large-scale domains, high dimensionality, or multiscale dynamics. 


The scalar advection-diffusion equation provides a canonical building block in conventional numerical methods development for computational fluid dynamics \cite{LeVeque2002}. It encapsulates two fundamental mechanisms of transport, advection, associated with bulk motion, and diffusion, representing molecular or turbulent mixing \cite{Batchelor2000}. Despite not completely covering all complexities such as pressure-velocity coupling and nonlinearities expressible with the complete Navier-St{\"o}kes set of governing equations, the scalar advection-diffusion equation presents a model analytical framework to test and benchmark early algorithms. It captures both the hyperbolic and parabolic behaviours in a single framework, providing for a robust and amenable prototype problem class. Quantum algorithms for time-dependent problems governed by PDEs can generally be categorized into two categories:  a) the variational quantum simulation methods \cite{demirdjian2022variational,leong2023variational} and b) Hamiltonian simulation techniques \cite{nielsen2010quantum}. The latter represents a more promising direction for long-term, fault-tolerant quantum computing \cite{ nielsen2010quantum,miyamoto2024quantum}. Hamiltonian simulation refers to the construction of a quantum circuit that models the time evolution of a quantum system, typically represented by the unitary operator $e^{-i\mathbf{H}t}$, where $\mathbf{H}$ denotes the system’s Hamiltonian and $t$ is the time increment. Studies have explored the use of Hamiltonian simulation by reformulating the governing equations of classical systems into an equivalent Schr\"odinger equation \cite{costa2019quantum, babbush2023exponential, jin2023quantum}.

Owing to inherent algorithmic and hardware constraints, computational fluid dynamics (CFD) presents a compelling opportunity for exploring quantum acceleration in classical engineering applications and remains an active area of investigation ~\cite{bharadwaj2020quantum,harrow2009quantum, yepez1998lattice, gourianov2022quantum,fukagata2022towards,oz2023efficient}. 
Steijl et al.~\cite{steijl2018parallel} demonstrated that a Poisson solver used in a vortex-in-cell approach could be reformulated as a hybrid quantum algorithm through the application of the quantum Fourier transform.
Gaitan~\cite{gaitan2020finding} proposed a quantum algorithm for solving the steady-state, inviscid, one-dimensional, compressible Navier–Stokes equations (NSE) in a converging-diverging nozzle, leveraging techniques from quantum ODE solvers~\cite{kacewicz2006almost}.
Extending this work, Oz et al.~\cite{oz2022solving} implemented a quantum algorithm to address the nonlinear Burgers equation.
In parallel, Suau et al.~\cite{suau2021practical} showcased a quantum approach for solving the wave equation based on the Hamiltonian simulation framework introduced by Costa et al.~\cite{costa2019quantum}. A recent study by Lee et al. ~\cite{lee2024multiple} proposed a multiple-circuit approach to reduce quantum resource requirements in the quantum lattice Boltzmann method, offering a practical direction for quantum computational fluid dynamics. Brearley et al.  used an explicit time marching strategy for solving the advection equation by embedding numerical integrators into a series of Hamiltonian simulations \cite{brearley2024quantum}. 
While these findings indicate that quantum algorithms have the potential to achieve speedup in simulating classical systems, they either typically assume oracle access to the Hamiltonian, or simulate idealistic models\cite{sato_central_difference} or may not apply to a general class of PDEs \cite{brearley2024quantum}. 


With a goal of building a long term general purpose PDE solver, here we adopt the Hamiltonian simulation approach with Schrödingerisation. In order to move forward with realistic calculations,  we consider vortex-dominated flows, characterized by circular or nearly circular streamlines and are governed by the dynamics of vorticity, a fundamental quantity in fluid mechanics \cite{wendt2008computational, batchelor2000introduction}. Vorticity, defined as the curl of the spatially varying transport field, quantifies the local rotational behavior of fluid elements and provides insight into the structure and evolution of complex flow fields. In particular, we will consider a transport field with non-zero vorticity, where the direction of the vorticity vector denotes the axis of rotation, while its magnitude reflects the intensity of local fluid spin. Vorticity plays a critical role in the analysis and modeling of key fluid dynamic phenomena, including boundary layers, shear layers, wakes, and mixing regions \cite{batchelor2000introduction}.

Our key contributions in this work can be summarized as follows:
\begin{itemize}
    \item{\textbf{Algorithmic scaling upto 30 qubits of Hamiltonian simulation of non-trivial transport fields}: We present a new algorithm for solving the 2D/3D advection-diffusion equation with spatially varying transport fields using upwinding discretization scheme for the advective component and central difference scheme for the diffusive component that scales to 30 qubits on emulator.}
    
    \item{\textbf{First hardware implementation up to 16 qubits}: Using the same discretization scheme, we implemented evolution of advection-diffusion equation in 2D for a (256x256 grid) with a spatially varying transport field for 40 time steps.}

    \item{\textbf{Efficient circuits for upwinding scheme}: We introduce practical approximations to the upwinding scheme for the advective component that eliminates the need for ancilla qubits, thereby removing mid-circuit measurements and reducing the overall circuit depth by at least 30\%.  }
\end{itemize} 



\section{Theory and Methods}\label{sec:theoryandmethods}
In this section, we will describe the form of the advection-diffusion equation we want to solve and the Schrödingerisation technique applied in order to convert it into a Schrödinger-like equation. 


\subsection{Advection-Diffusion Equation and Schr\"odingerisation}
A $d$-dimensional advection-diffusion equation is written as,
\begin{align}
    \pdv{u(t,\mathbf{x})}{t} &= -v\cdot \grad{u(t,\mathbf{x})} + D\grad^2{u(t,\mathbf{x})}  \nonumber \\ &\equiv -\sum_{\alpha=1}^{d} v_{\alpha} \partial_{\alpha} u(t,\mathbf{x}) + D\grad^2{u(t,\mathbf{x})},
\label{eq:advection}
\end{align} 

where $u$ is the scalar field moving with transport $v$ in a medium with diffusivity $D\geq0$. $\grad$  and $\grad^2$ are the spatial gradient and diffusion operators respectively. Applying the warped
phase transformation $\mathbf{v}(t,p) = e^{-p}u(t,\mathbf{x})$, 
and subsequently applying a Fourier transform ($\mathcal{F}$) over $p$, ($\hat{\mathbf{v}(t,\eta)} = \mathcal{F}(\mathbf{v}(t,p))$), a spatially discretized version of Eq.~\eqref{eq:advection} can  be written as 
\begin{align}
    \pdv{\hat{\mathbf{v}}(t,\mathbf{x},\eta)}{t} = \mathbf{H_D}\hat{\mathbf{v}}(t,\mathbf{x},\eta)+\mathbf{H}\hat{\mathbf{v}}(t,\mathbf{x},\eta),
\label{eq:schrodingerised_equation}
\end{align}
where we refer to $\mathbf{H_D}$ as the diffusion Hamiltonian and $\mathbf{H}$ as the advection Hamiltonian.

The discretization process  has been described in \cite{Hu2024quantumcircuits} with respect to heat and advection equations, such that the $\mathbf{H_D}$ comes from the diffusion term in the heat equation and $\mathbf{H}$ comes from the gradient term of advection equation. For the ease of readers, we describe the process briefly here. We first discretize the variable $p$ over $N_p = 2^{n_p}$ grid points in the warped space $[-\pi R, \pi R]$ (for $R \in \mathbb{R}$) and the variable $x_{\alpha}$ over $N_{\alpha} = 2^n_{\alpha}$ grid points on the physical space $[0,L_{\alpha}]$ such that the grid size is $h_{\alpha}=L_{\alpha}/N_{\alpha}$. For simplicity, we choose all $N_{\alpha}$ and $L_\alpha (=  L))$ to be the same and hence call all $h_{\alpha}$ to be $h=1$. The discretized scalar field $u(t,\mathbf{x})$ is encoded in the quantum state $\ket{u(t,\mathbf{x})}$ using amplitude encoding such that 
\begin{align}
    \ket{u(t,\mathbf{x})} = \sum_{j_1=0}^{L-1 }\ldots \sum_{j_d=0}^{L-1 } u(t,x_{j_1}\ldots ,x_{j_d})\ket{j_1}\otimes \ldots \otimes \ket{j_d},
    \label{eq:encoding}
\end{align}
where $u$ is a normalized array and $\ket{j_i}$ is the computational basis for the $x_i$ axis. We further define the Fourier space variable $\eta_k = (k-N_p/2)$, $k= 0,1,\ldots,N_p-1$. 
Using all the discretizing ingredients, the Hamiltonians  $\mathbf{H_D}$ and $\mathbf{H}$ are given by,
\begin{align}
&\mathbf{H_D} = \sum_{k=0}^{N_p} \qty( k-\frac{N_p}{2} ) \sum_{\alpha=1}^{d} \mathbf{H}_{D,\alpha} \otimes \dyad{k}{k}  \nonumber \\
&\mathbf{ H} = \sum_{k=0}^{N_p} \qty( k-\frac{N_p}{2} ) \sum_{\alpha=1}^{d} \mathbf{H}_{1,\alpha} \otimes \dyad{k}{k} + \sum_{\alpha=1}^{d} \mathbf{H}_{2,\alpha} \otimes I^{\otimes N_p}.
\label{eq:full_hamiltonian},
\end{align}
where it should be noted that we use the same set of ancilla qubits for both $\mathbf{H_D}$ and $\mathbf{H}_{1,\alpha}$. In 3D we have $\alpha\in \{x,y,z\} $. Following \cite{Hu2024quantumcircuits,sato_central_difference}, we can expand $\mathbf{H}_{D,\alpha}$, $\mathbf{H}_{1,\alpha}$ and $\mathbf{H}_{2,\alpha}$  as,
\begin{align}
\mathbf{H}_{D,\alpha} & =  \gamma_D^{\alpha} \left(\qty( \sigma_{01}^{\otimes n_{\alpha}} + \sigma_{10}^{\otimes n_{\alpha}}) + \sum_{j=1}^{n_{\alpha}}\qty( s_j^{-} + s_j^{+}) -2 I^{\otimes n_{ \alpha}}\right) \nonumber   \\  &\equiv D S_{D,\alpha} \label{eq:s3} \\
\mathbf{H}_{1,\alpha} & =  \gamma_1^{\alpha} \left(\qty( \sigma_{01}^{\otimes n_{\alpha}} + \sigma_{10}^{\otimes n_{\alpha}}) + \sum_{j=1}^{n_{\alpha}}\qty( s_j^{-} + s_j^{+}) -2 I^{\otimes n_{ \alpha}}\right) \nonumber   \\  &\equiv |v_{\alpha}| S_{1,\alpha} \label{eq:s1}\\
\mathbf{H}_{2,\alpha} &= -i\gamma_2^{\alpha} \left( \qty( -\sigma_{01}^{\otimes n_{\alpha}} + \sigma_{10}^{\otimes n_{\alpha}}) + \sum_{j=1}^{n_{\alpha}}\qty( s_j^{-} - s_j^{+})\right) \nonumber \\  &\equiv v_{\alpha} S_{2,\alpha}. \label{eq:s2} 
\end{align}
where we have separated out the $\gamma$'s so that we can build our theory on top of the theory of the constant transport model for the advection Hamiltonian. For later convenience, the total Hamiltonian for a given physical dimension labeled with $\alpha \in \{x,y,z\}$ is defined as 
\begin{align*}
    \mathbf{H_{D,\alpha}}+\mathbf{H_{1,\alpha}}+\mathbf{H_{2,\alpha}} = D S_{D,\alpha}+ |v_{\alpha}|S_{1,\alpha}+v_{\alpha}S_{2,\alpha}.
\end{align*}
Writing the $\gamma$'s explicitly,
\begin{align}
\gamma_{D}^{\alpha} = \frac{D}{h^2R}, \gamma_{1}^{\alpha} = \frac{|v_\alpha|}{2hR} , \gamma_{2}^{\alpha} = \frac{v_\alpha}{2h}.
\end{align}
In the above equations, we have used  $2\times2$ matrices,
\bea 
\sigma_{01} &=& \dyad{0}{1};\;\sigma_{10} = \dyad{1}{0}  \\
\sigma_{00} &=& \dyad{0}{0};\;\sigma_{11} = \dyad{1}{1}
\eea 
and 
\bea
s_j^{+} &=& I^{\otimes(n_x - j)}\otimes \sigma_{10} \otimes \sigma_{01}^{\otimes(j)} \\
s_j^{-} &=& I^{\otimes(n_x - j)}\otimes \sigma_{01} \otimes \sigma_{10}^{\otimes(j)}
\eea

\subsection{Circuit Implementation}
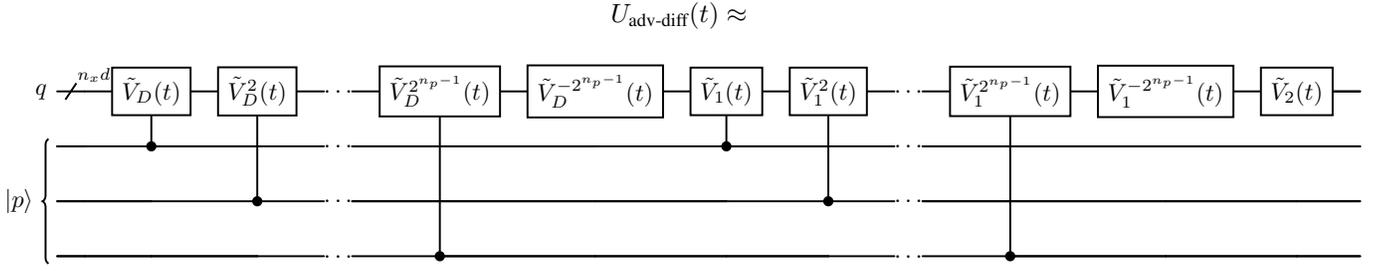
\begin{figure*}[htbp]
    \centering
      $U_{\text{adv-diff}}(t)  \approx$ 
        \begin{adjustbox}{width=\textwidth,center} 
    \begin{quantikz}[row sep={0.8cm,between origins},column sep=0.4cm,
        remember picture]
         \\ 
         \lstick{$q$}  & \qwbundle{n_x d} & \gate{\tilde{V}_{D}(t)} &           \gate{\tilde{V}_{D}^{2}(t)}  & \ldots & \gate{\tilde{V}_{D}^{2^{n_p-1}}(t)}  & \gate{\tilde{V}_{D}^{-2^{n_p-1}}(t)}  & \gate{\tilde{V}_{1}(t)} &           \gate{\tilde{V}_{1}^{2}(t)} & \ldots & \gate{\tilde{V}_{1}^{2^{n_p-1}}(t)} & \gate{\tilde{V}_{1}^{-2^{n_p-1}}(t)} &                           \gate{\tilde{V}_{2}(t)} & \qw \\  
        \lstick[wires=3]{$\ket{p}$}
         & \qw & \ctrl{-1} & \qw  & \ldots & \qw  & \qw & \ctrl{-1} & \qw  & \ldots & \qw & \qw & \qw  & \qw \\         
          & \qw & \qw & \ctrl{-2}  & \ldots & \qw  & \qw  & \qw &   \ctrl{-2}   & \ldots & \qw & \qw & \qw & \qw  \\  
         & \qw & \qw & \qw  & \ldots & \ctrl{-3} & \qw & \qw  & \qw &  \ldots &  \ctrl{-3}  & \qw  & \qw & \qw  \\         
    \end{quantikz}
      \end{adjustbox} 
    \caption{Quantum circuit for $U_{\text{adv-diff}}(t)$ for a constant (zero vorticity) transport advection evolution. } 
    \label{fig:adv:circuit:V}
\end{figure*}

The full unitary for the advection-diffusion evolution $U_{\text{adv-diff}}(t)$ is implemented via three Hamiltonian operators $\mathbf{H_{D,\alpha}}, \mathbf{H_{1,\alpha}}$, and $\mathbf{H_{2,\alpha}}$. This unitary can be approximated as
\begin{equation}\label{eqn:adv:gate:Vtau}
\begin{aligned}
    U_{\text{adv-diff}}(t)  & = \exp{i(\mathbf{H_{D,\alpha}}+\mathbf{H_{1,\alpha}}+\mathbf{H_{2,\alpha}})dt}\nonumber \\& \approx \exp(i\mathbf{H}_{2,\alpha}dt)\exp(i\mathbf{H}_{1,\alpha}dt)\exp(i\mathbf{H}_{D,\alpha}dt) \nonumber \\&\approx 
    \left( \tilde{V}_{2}(t) \otimes I^{\otimes n_p} \right)  \times\left(\sum_{k=0}^{N_p- 1} \tilde{V}_{1}^{k-N_p/2}(t) \otimes |k\rangle \langle k|\right)\\&\times\left(\sum_{k=0}^{N_p- 1} \tilde{V}_{D}^{k-N_p/2}(t) \otimes |k\rangle \langle k|\right),
\end{aligned} 
\end{equation}

where we can obtain further simplification by 
\begin{align}
    \sum_{k=0}^{N_p- 1} \tilde{V}_{1}^{k}(t) \otimes |k\rangle \langle k| = \sum_{m=0}^{n_p- 1} \tilde{V}_{1}^{2^m}(t)\otimes\ket{1}\bra{1}+\mathcal{I}^{n_{\alpha}}\otimes\ket{0}\bra{0} \label{Eq:V1component}. \\
    \sum_{k=0}^{N_p- 1} \tilde{V}_{D}^{k}(t) \otimes |k\rangle \langle k| = \sum_{m=0}^{n_p- 1} \tilde{V}_{D}^{2^m}(t)\otimes\ket{1}\bra{1}+\mathcal{I}^{n_{\alpha}}\otimes\ket{0}\bra{0}.\label{Eq:diffusion_component}
\end{align}
The resulting full circuit is given in Fig.~\ref{fig:adv:circuit:V} where the gates $\tilde{V}_{D}(t)$, $\tilde{V}_{1}(t)$, and $ \tilde{V}_{2}(t)$ are defined as,
\bea 
\tilde{V}_{D}(t) &\approx& \prod_{\alpha=1}^{d}  \exp{-i\mathbf{H}_{D,\alpha}t}, \label{eq:v3}\\
\tilde{V}_{1}(t) &\approx& \prod_{\alpha=1}^{d}  \exp{-i\mathbf{H}_{1,\alpha}t}, \label{eq:v1}\\
\tilde{V}_{2}(t) &\approx& \prod_{\alpha=1}^{d} \exp{-i\mathbf{H}_{2,\alpha}t} \label{eq:v2}
\eea 
Circuit implementation for $\tilde{V}_{D}(t)$, $\tilde{V}_{1}(t)$ and $\tilde{V}_{2}(t)$ are followed from \cite{ezzell2023} and have been defined in the Appendix \ref{appendix:circuitsv1v2}.

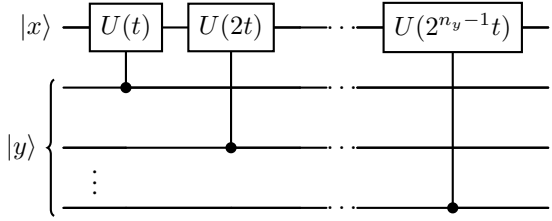
\begin{figure}[t]
\begin{quantikz}[row sep={0.8cm,between origins}, column sep=0.35cm]
\lstick{$\ket{x}$} &  \gate{U(t)}  \qw & \gate{U(2t)} & \qw & \ldots & \gate{U(2^{n_{y}-1}t)} \qw & \qw\\
\lstick[wires=3]{$\ket{y}$}
& \ctrl{-1} & \qw & \qw & \ldots & \qw & \qw\\
& \qw{\vdots} &\ctrl{-2} & \qw & \ldots  & \qw &  \qw  \\
& \qw &\qw  & \qw &  \ldots &\ctrl{-3} & \qw \\
\end{quantikz}

\caption{Modified circuit for incorporating a linearly varying transport field $v=(y,0)$. $U(2^jt)$ is the unitary that implements advection with transport $v_{\alpha}=2^j$.}
\label{fig:adv_controlled_y}
\end{figure}

\subsection{Spatially varying transport field}\label{sec_variable_vel}
As mentioned in the introduction, we will consider the advection-diffusion equation with a spatially varying transport field. Notice that the diffusion component of the circuit plays no role here. 
 To accurately capture such behavior in CFD models, it is essential to incorporate transport fields that preserve vorticity into the governing PDEs.
Existing works on solving advection equations with Hamiltonian simulation have only explored a uniform transporting field ($v_x=v_y=c$) \cite{Hu2024quantumcircuits,sato_central_difference, PhysRevA.110.012430}. In this work, 
 we will consider an implementation of a non-trivial transport field with spatial dependence.  
 We will denote the qubit registers for $x$ and $y$ directions as $\ket{x}$ and $\ket{y}$, respectively and the scalar field would be represented as
 \be
 \ket{u(x,y,t)} = \sum_{x,y}c_{x,y,t} \ket{x}\ket{y}.
 \label{eq:statevector}
\ee

 
 We consider the transport field $v = (y,x)$ and work in the first quadrant $y\in [0,L]$ such that $v_x\geq0$. Therefore, the advection Hamiltonian has the following form \begin{align}
     \mathbf{H} &= \mathbf{H_{x}}+\mathbf{H_{y}} \nonumber\\
     &= y(S_{1,x}+S_{2,x})+x(S_{1,y}+S_{2,y}) \\&= yS_x+xS_y, \nonumber
 \end{align}  
where  $S_{x}=S_{1,x}+S_{2,x}$ and $S_{y}=S_{1,y}+S_{2,y}$. To demonstrate how this Hamiltonian can be simulated in the upwinding scheme, we focus on the implementation of the $x$-component of the Hamiltonian $\mathbf{H_{x}}$. The implementation for $\mathbf{H_{y}}$ can be done in a similar way. Let us denote the operator $U(t) = e^{-iS_{x} t}$, for later convenience. Note that this is the evolution operator for the advection equation with transport $v_x=1$, the implementation of which is already described in \cite{Hu2024quantumcircuits}. 
Using the fact that $y = \sum_{m=0}^{n_x-1} y_m 2^m$, where $y_m \in \{0,1\}$, it can now be shown that 

\begin{align}
e^{-i\mathbf{H_{x}}t} \ket{x} &= \sum_{y=0}^{N_{y}-1}\left(e^{-i yS_{x} t}\otimes \ket{y}\bra{y}\right) \ket{x} \nonumber \\
&= \sum_{y=0}^{N_{y}-1}\left(e^{-i(\sum_{m=1}^{n_y} y_m 2^m) S_{x}t } \otimes \ket{y}\bra{y}\right) \ket{x} \nonumber \\
&= \left(\mathop{\prod\nolimits'}_{m=0}^{n_y-1} e^{-i y_m 2^m S_xt }\otimes \sum_{y_m}\ket{y_m}\bra{y_m}\right)\ket{x} \nonumber \\ & = \left(\mathop{\prod\nolimits'}_{m=0}^{n_y-1}U(y_m2^mt)\otimes \sum_{y_m}\ket{y_m}\bra{y_m}\right)\ket{x} \nonumber \\
& = \left(\mathop{\prod\nolimits'}_{m=0}^{n_y-1}U(2^mt)\otimes \ket{1}\bra{1}+\mathcal{I}^{n_{x}}\otimes\ket{0}\bra{0}\right)\ket{x},\label{Eq:var_vel_algo}
\end{align}

where the primed product $\mathop{\prod\nolimits'}$ implies direct product with respect to $\ket{x}$ register and tensor product with the $\ket{y}$ register. The above equation can be implemented using the circuit in Fig.~\ref{fig:adv_controlled_y}, where the unitary $U(t)$ is being controlled by $\ket{y}$ qubits. It should be noted that $ U_{\text{adv}}(t) \approx U(t)$ for $v_x=1$. The above derivation can be extended to transport fields, where $v_x=-y$, i.e., with a negative and linearly varying transport field. In this case, 
\begin{align*}
    \mathbf{H_x} = y S_{1,x}-yS_{2,x} = yS_{x}',
\end{align*}
where $S_{x}'= S_{1,x}-S_{2,x}$. By defining $U'(t)= e^{-iS_x't}$, we can simply follow the previous derivation to obtain the corresponding circuit.

\subsection{Gate complexity of the modified circuit}
In Ref. \cite{Hu2024quantumcircuits}, complexity analysis for the heat equation and advection equation with uniform transport field was done. Our algorithm is built on top of their method by observing that the heat equation operator is essentially the diffusion operator and appending it with the modified advection equation operator for the spatially varying transport field. Therefore, the gate complexity of the diffusion component Eq.~\eqref{Eq:diffusion_component} of our algorithm is the same as the heat equation from \cite{Hu2024quantumcircuits}, which states that for $n_{\alpha} \geq 3$

\begin{align}\label{Eq:Diff_gate_complexity}
    1-qubit_{Diff} = \mathcal{O}(N_pn_{\alpha}), \nonumber\\
    CNOT_{Diff} = \mathcal{O}(N_pn^2_{\alpha}),
\end{align}
where $1-qubit_{Diff}$ and $CNOT_{Diff}$ represent the single qubit and CNOT gate counts respectively for the diffusion component. For $d$-dimensions the corresponding gate count is simply multiplied with $d$. Now, the advection component of the circuit that consists of $\tilde{V}_2(t)\otimes I^{\otimes n_p}$ and Eq.~\eqref{Eq:V1component}. For one-dimension, with $n_{\alpha}\geq 3$ the gate complexity is given by 

\begin{align}
    {1-qubit}_{Adv} &= 6n_{\alpha}+2+2^{n_p-1}(4_{\alpha}+1)=\mathcal{O}(N_pn_{\alpha}), \nonumber \\
    {CNOT}_{Adv} &= (2^{n_p}-1)(16n_{\alpha}^2-2n_{\alpha}-30) \nonumber \\& + (2^{n_p-1}+1)9n_{\alpha}^2-15n_{\alpha}-8 = \mathcal{O}(N_pn_{\alpha}^2),
    \label{Eq:old_Adv_gate_complexity}
\end{align}
where ${1-qubit}_{Adv}$ and ${CNOT}_{Adv}$ represent the single qubit and CNOT gate count respectively. For constant transport fields, if there are $d$-dimensions the corresponding number is obtained by multiplying with $d$. For spatially varying transport fields, let's consider $d=1$, since in this case a simple multiplication with $d$ won't give the correct gate count. For the modified circuit in Fig.~\ref{fig:adv_controlled_y}  the corresponding CNOT gate count is given by 
\begin{align}
    Adv_{CNOT} &= (2^{n_\alpha}-1)(\mathcal{O}(N_pn_{\alpha})+Adv_{CNOT}*8)  \nonumber \\ &= \mathcal{O}(N_pN_{\alpha}n_{\alpha}^2) \approx \mathcal{O}(N_pN_{\alpha}).
    \label{Eq:Adv_gate_complexity}
\end{align}

Therefore the CNOT gate count is now exponential with respect to both the system qubits $n_{\alpha}$ and ancilla qubits $n_p$. For the full circuit in Fig.\ref{fig:adv:circuit:V} and for a given dimension $\alpha$, the corresponding gate complexity is obtained by adding Eq.~\eqref{Eq:Diff_gate_complexity} and Eq.~\eqref{Eq:Adv_gate_complexity}, such that 
\begin{align*}
    &1_{qubit} = \mathcal{O}(N_pn_{\alpha}), \\
    &CNOT = \mathcal{O}(N_pn^2_{\alpha})+\mathcal{O}(N_pN_{\alpha}) \approx \mathcal{O}(N_pN_{\alpha}).  
\end{align*}

\subsection{Efficient circuit implementation without ancillae}
\label{sec:circuit_approx}
To be able to run our algorithm on currently available quantum hardware that have constraints on the qubit count and circuit depth, we need to make certain approximations to the advective component of the circuit.  We will ignore the $c-V_1$ component of the circuit from Fig~\ref{fig:adv:circuit:V}. This approximation is justified due the following to two reasons. 

Following \cite{Hu2024quantumcircuits}, the advection Hamiltonian can be written as a sum of the finite difference operators, $\mathbf{H} = \eta A_1 + A_2$ with
\begin{align*}
    A_1 = \sum_{\alpha=1}^d \frac{|v_{\alpha}|h}{2}(D^{\Delta}_P)_{\alpha}, \quad 
    A_2 = \sum_{\alpha=1}^d \frac{v_{\alpha}}{i}(D^{\pm}_P)_{\alpha}. 
\end{align*}
where $(D_P^{\Delta})_{\alpha}$ and $(D_P^{\pm})_{\alpha}$ represent the second-order and first-order derivatives using the central finite difference respectively.

The primary phenomenon of advection is implemented by the $A_2$ component which is the first-order derivative using the central difference scheme. Advection with $A_2$ alone is prone to numerical oscillations \cite{molenkamp1968accuracy,van1979towards}, which is stabilized by the second-order diffusion term $A_1$.  Evolutions due to $\mathbf{H}_{1,\alpha}$'s are approximated by $A_1$ and implemented using the quantum gate $V_1$ Eq.~\eqref{eq:v1}. 
To compare the contribution of the $A_1$ compared to $A_2$, consider the following 
\begin{align}
    A_{1/2}=\frac{|A_1|}{|A_2|} = \frac{h|D^{\Delta}_P|}{2|D^{\pm}_P|} \approx h\frac{|u''(x)|}{|u'(x)|}.
\end{align}
$A_{1/2}$ is small when we have a finer grid, i.e. $h$ is small, and also when the ratio of second-order derivative to first-order derivative is small. In our case, the grid size $h$ is always 1, so it does not affect $A_{1/2}$. However, since we have a superposition of Gaussian states, this ratio is inversely proportional to the grid size, i.e. $A_{1/2}\propto \frac{1}{L^2}$. This happens because the $\sigma$ is proportional to grid size $L$. Therefore, as the size of the grid increases, the ratio $A_{1/2}$ becomes increasingly small. Thus, any approximations that we make to the $V_1$ component (The block in Fig.~\ref{fig:adv:circuit:V} containing only the unitary gate $V_1$) of the circuit will not have a significant effect on the final output for a sufficiently large grid size.

Second, the only input to $V_1$ is $2\gamma_1 dt$.  If this input is zero $V_1$ becomes identity, thus, if we make $dt$ sufficiently small $c-V_1$ (controlled-$V_1$) will become much closer to identity. Therefore, we calculate the distance of the operator $\mathcal{U}_{cV1}$ (corresponding to the circuit $c-V_1$) from identity using the normalized Frobenius norm given by
\begin{align*}
    D(\mathcal{U}_{cV1},\mathcal{I})= \frac{\|\mathcal{U}_{cV1}-I\|_F}{\sqrt{L_{}}}=\frac{\sqrt{\sum_{i}^{L_{}}\sum_{j}^{L_{}}|\mathcal{U}_{ij}-\delta{ij}|^2}}{\sqrt{L_{}}}.
\end{align*}

This distance is obtained for $n_x=8$, $R=8$, $d=1$, and $dt=0.1$ for which we have done the hardware run. For $n_p=1,2$, the distance is of the order of $1e^{-2}$, whereas for a 2D problem, i.e., for $d=2$, the corresponding distance is of the order $1e^{-3}$. Thus, to a good degree of approximation, we can assume that the $c-V_1$ component of the circuit is close to the identity and therefore can be ignored for our hardware implementation. Since the full quantum circuit for a single time step is made from three components, $c-V_1, V_1^{-1}$ and $V_2$, by ignoring $c-V_1$ at least one-third of the circuit depth can be reduced. In general, these approximations will hold true for smoothly varying functions where the second order is much smaller than the gradient.

\begin{figure*}[t]
    \includegraphics[width=\linewidth]{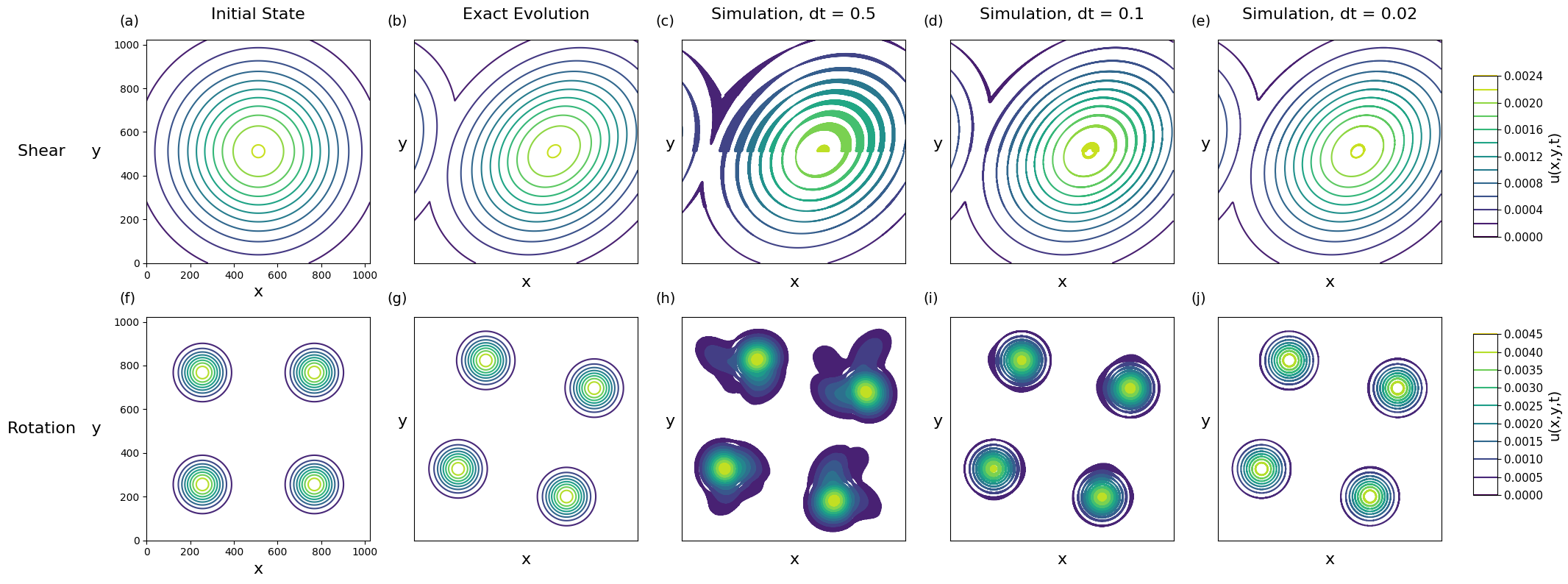}    \caption{\textbf{Simulation results of the advection equation with periodic BC in a $\mathbf{1024\times 1024}$ (20 qubits) lattice}: The top row represents the evolution of the scalar field $u_{shear}(x,y,t)$ with transport field $(\frac{y}{L/2},0)$. Plot (a) represents the initial state while (b) is the exact analytical solution at time $T=128$. Plots (c), (d), and (e) are the solutions obtained from the proposed method at time $T=128$, with $dt=0.5,0.1,$ and $0.02$ respectively.  Similarly, the bottom row represents the evolution of the scalar field $u_{rotation}(x,y,t)$ with transport field $(\frac{y-L/2}{L/2},\frac{-(x-L/2)}{L/2})$. For the initial state is given in (f), the analytical solution is plotted in (g) at time $T=128$. Similar to shear, we have plotted the solutions from the quantum simulation for $dt=0.5, 0.1,$ and $0.02$ in (h), (i), and (j) respectively. In both scenarios, we observe improvement in the numerical solution with a decrease in time increment size $dt$.} 
    \label{fig:sim_evolution}
\end{figure*}

\section{Simulation Results}
 
\subsection{Simulation - Advection Equation}
We first, present simulation results for the advection equation to confirm the expected advective transport behavior. For classical simulation, we have used CUDA-Q 0.9.1 \cite{cudaq}, an open-source toolkit that can be used for building and executing quantum circuits on a local simulator. As the backend, the \textit{Statevector} simulator was used to get the full quantum state. 

\subsubsection{Two dimensional Shear}\label{subsubsection:shear}
First, we consider the two-dimensional advection equation with periodic boundary condition  (PBC) and transport field $v = (\frac{y}{L/2},0 )$, such that the Courant–Friedrichs–Lewy  (CFL) condition is maintained for numerical stability. The advection equation for shear in 2D with PBC is written as,
\begin{align}
\begin{cases}\label{Eq:shear_advection}
    \frac{\partial u(x,y,t)}{\partial t} + \frac{y}{L/2}\frac{\partial u(x,y,t)}{\partial x} = 0,  \\
    u(x,y,t) = u(x+L,y,t),  \quad u(x,y,t) = u(x,y+L,t).\\
\end{cases}
\end{align}

where  as $x, y \in [0, L]$. The initial state $u_{shear}(x,y,0)$ is a 2D Gaussian, 
\begin{align}
    u_{shear}(x,y,0) = \frac{1}{\mathcal{N}}\exp\left(-\frac{(x-\mu)^2+(y-\mu)^2}{2\sigma^2}\right),
    \label{Eq:initial_state_shear}
\end{align}
where  $\mu = L/2$, $\sigma = L/4$, and $\mathcal{N}$ is the normalization factor. This is chosen so as to have a smooth initial state such that we can get very good accuracy with just one ancilla. If we let this evolve according to Eq.~\eqref{Eq:shear_advection} for time $T$, the analytical solution is given by 
\begin{align}
    &u_{shear}(x,y,T)  = \frac{1}{\mathcal{N}}\exp\left(-\frac{(x-\frac{y}{L/2}T-\mu)^2+(y-\mu)^2}{2\sigma^2}\right).  
    \label{Eq:analytical_shear}
\end{align}

Under the action of this transport field, different layers of the fluid move along the $x$-direction with velocities that vary linearly with respect to the $y$-direction. This can be implemented using the circuit from Eq.~\eqref{Eq:var_vel_algo} with the only difference being that the transport field $v_x=y$ is normalized by dividing with $L/2$.

In Fig. \ref{fig:sim_evolution}(a-e), we show the simulation results of the evolution of the 2D scalar function Eq.~\eqref{Eq:initial_state_shear} for $n_x=n_y=10$, i.e, $1024\times 1024 (L\times L)$ grid points.  We have chosen  $R=4$ and $T=128$ and run the simulation for different time stepsize $dt=0.5$, $dt=0.1$, $dt=0.02$. It can be seen that by decreasing $dt$, the proposed method approaches the analytical solution, thereby validating our proposed method. 
For various values of $dt$, we also calculate the relative $\ell_2$-norm $\mathbf{e}_{dt}$, defined as
\begin{align}
    \mathbf{e}_{dt}=\frac{\|u_{shear}(x,y,T)-u_Q(x,y,T)\|_2}{\|u_{shear}(x,y,T)\|_2}
    \label{Eq:l2norm}
\end{align}
where $u_Q(x,y,T)$ is the numerical solution obtained by the proposed algorithm. Here the denominator is always 1 since $u_{shear}(x,y,T)$ is normalized.  For the different cases plotted in Fig.~\ref{fig:sim_evolution}(c-e), the $\ell_2$-norm errors are $\mathbf{e}_{0.5}=2.88e^{-2}$, $\mathbf{e}_{0.1}=5.89e^{-3}$ and $\mathbf{e}_{0.02}=2.25e^{-3}$.


\subsubsection{Two dimensional Rotation}\label{subsubsection:rotation}
We consider a 2D advection equation with periodic boundary conditions, and with a transport field  $v = (\frac{y-L/2}{L/2},\frac{-(x-L/2)}{L/2})$. The advection equation can be put together as
\begin{align}
\begin{cases}\label{Eq:rotation_advection}
    \frac{\partial u(x,y,t)}{\partial t} + \frac{y-L/2}{L/2}\frac{\partial u(x,y,t)}{\partial x}-\frac{(x-L/2)}{L/2}\frac{\partial u(x,y,t)}{\partial y} = 0, \\
    u(x,y,t) = u(x+L,y,t),  \quad u(x,y,t) = u(x,y+L,t).
\end{cases}
\end{align}
where  $x,y \in [0, L]$. For the initial state $u_{rotation}(x,y,0)$, we choose a superposition of 2D Gaussian states 
\begin{align}
   u_{rotation}(x,y,0) = \sum_{i,j=\{1,2\}}\frac{1}{\mathcal{N}}\exp\left(-\frac{(x-\mu_i)^2+(y-\mu_j)^2}{2\sigma^2}\right),
   \label{Eq:initial_state_rotation}
\end{align}
where the constants have values $\mu_1 = L/2+L/4$, $\mu_2 = L/2-L/4$,  $\sigma = L/16$, and $\mathcal{N}$ is the normalization factor. The choice of initial state helps us to observe the evolution implemented with this transport field which is a clockwise rotation around the point $(L/2, L/2)$. Generally, a clockwise rotation around $(0,0)$ is implemented via the transport field of the form $v=(y,-x)$. This leads to a rotation around a corner in our implementation. In order to see the effect of the rotation more prominently, we shift the center for rotation to $(L/2,L/2)$.  The analytical solution to this problem at time $T$ is given by,
\begin{align}
    &u_{rotation}(x,y,T) \nonumber \\ &= \frac{1}{\mathcal{N}}\sum_{i,j =\{1,2\}}\exp\left(-\frac{(x'(T)-\mu_i)^2+(y'(T)-\mu_j)^2}{2\sigma^2}\right),    
    \label{Eq:analytical_rotation}
\end{align}
where the rotated coordinates are 
\begin{align*}
    &x'(T)=(x-\frac{L}{2})\cos(\frac{2T}{L})-(y-\frac{L}{2})\sin(\frac{2T}{L})+\frac{L}{2}+mL,\\
    &y'(T)=(x-\frac{L}{2})\sin(\frac{2T}{L})+(y-\frac{L}{2})\cos(\frac{2T}{L})+\frac{L}{2}+m'L,    
\end{align*} 
where $m, m' \in \mathbb{Z}$ . Implementing this shifted rotation with our method is a two-step application of Eq.~\eqref{Eq:var_vel_algo} for every time step. The first step is the advection along $x-$direction with a transport $\frac{y}{L/2}$ followed by an advection with constant transport $-1$, which makes the resulting $x$ velocities to be $\frac{y-L/2}{L/2}$. Similarly, the second component is an advection along the $y-$direction with transport $\frac{-x}{L/2}$  followed by an advection with constant transport $1$, which results in $v_y=\frac{L/2-x}{L/2}$.

The corresponding numerical solution has been plotted in Fig.~\ref{fig:sim_evolution} (f-j). We choose the same problem size as the shear, i.e, for $n_x=n_y=10$, i.e, $1024\times 1024 (L\times L)$ grid points and $R=4$ and $T=128$. We have run the simulation for different time step sizes $dt=0.5$, $dt=0.1$, $dt=0.02$. 
Again we observe that the results agrees closely with the analytical solution for $dt=0.02$. Since the source of error in our proposed method is due to Trotterization, it can be improved by reducing the time step size, which is verified in Fig.~\ref{fig:sim_evolution}(h-j), where we see the improvement in simulation results with decreasing $dt$. In this case the $\ell_2$-norm of relative errors for different values of $dt$ are  
$\mathbf{e}_{0.5}=0.32$, $\mathbf{e}_{0.1}=6.88e^{-2}$ and $\mathbf{e}_{0.02}=1.89e^{-2}$.

\subsection{Simulation- Advection-Diffusion Equation}
\subsubsection{Two-dimensional Shear and Rotation}
\begin{figure}[t]
    \includegraphics[width=\linewidth]{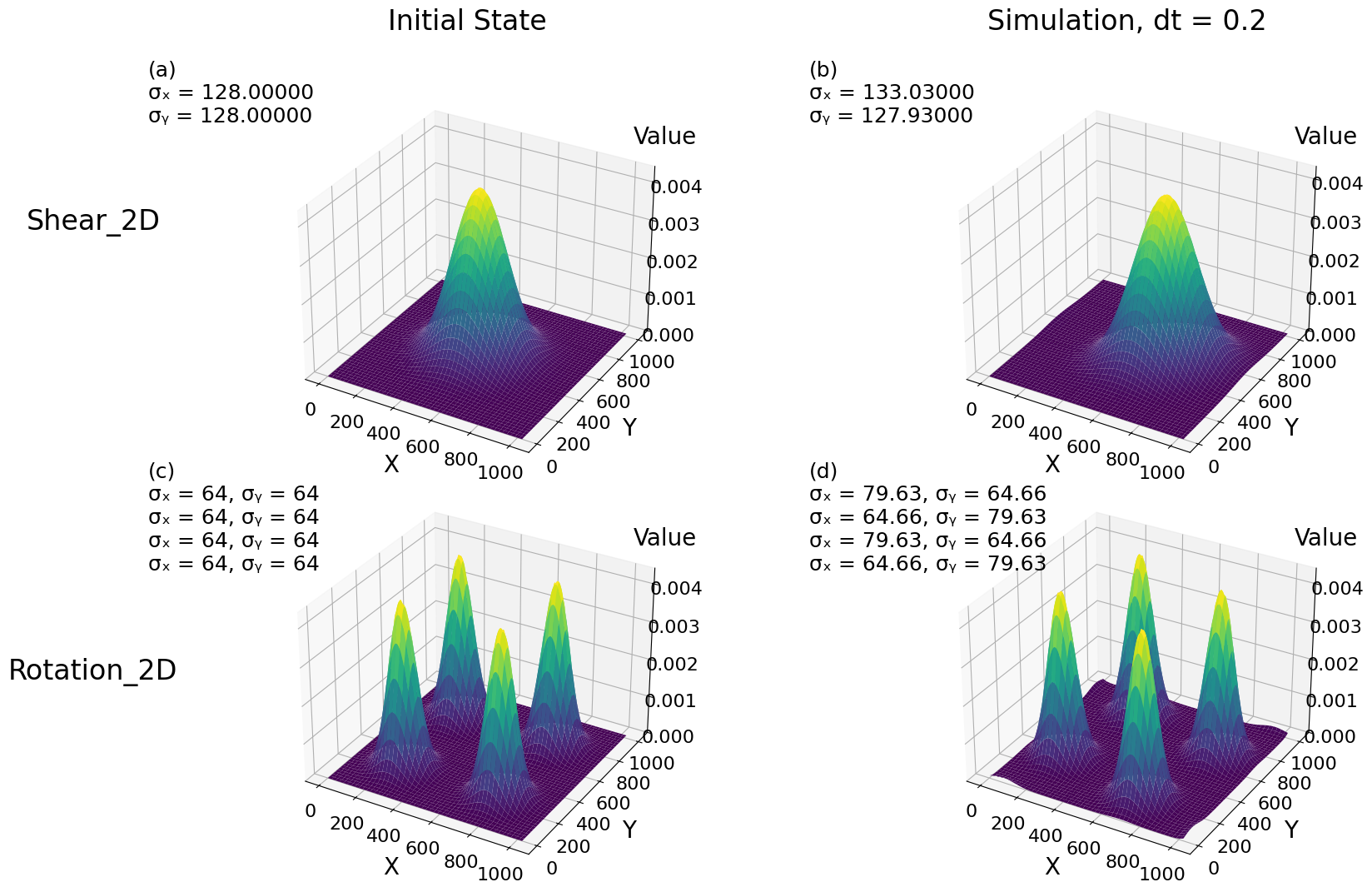}    \caption{\textbf{Simulation results of the advection-diffusion equation with periodic BC in a $\mathbf{1024\times 1024}$  lattice}: The initial states for shear and rotation are plotted in (a) and (c). We run both simulations for a time $T=128$ with $dt = 0.2$. The final form of the evolved states is plotted in (b) and (d). It is readily seen that the evolved gaussian are more spread out due to diffusion. For brevity, we have mentioned the variance values for the corresponding plots, which tell about the degree of spread due to diffusion.} 
    \label{fig:adv_diff2D}
\end{figure}

So far, we have presented the simulation results for the advection equation only. In this section, we will present the simulation results by including the diffusion term as well. The advection-diffusion equations for the two-dimensional shear and rotation transport fields are obtained by adding a diffusion term to the right-hand side of the equations Eq.~\eqref{Eq:shear_advection} and  Eq.~\eqref{Eq:rotation_advection}, respectively. The 2D diffusion term is of the following form 

\begin{align}\label{Eq:diffusion_terms}
    &\nabla^2u(x,y,t) = D\left(\frac{\partial^2 u(x,y,t)}{\partial x^2} + \frac{\partial^2 u(x,y,t)}{\partial y^2}\right),
\end{align}
where the diffusion coefficient has the value $D=1$ for all the results presented in this section. We take the same initial Gaussian states for the respective cases as in Eq.~\eqref{Eq:initial_state_shear} and Eq.~\eqref{Eq:initial_state_rotation}. For the 2D shear transport field, we change the variance $\sigma = L/8$ due to the fact that diffusion terms becomes visibly significant for a peaked Gaussian initial state.  The corresponding analytical solutions in the presence of the diffusion term maintains the same mathematical form as in Eq.~\eqref{Eq:analytical_shear} and Eq.~\eqref{Eq:analytical_rotation}, with an enlarged variance, which at time $T$ is given by
\begin{align}\label{enlargedsigma}
    \sigma_T^2 = \sigma^2+2DT,
\end{align}
where $\sigma_T^2$ and $\sigma^2$ are the variances at time $T$ and at the beginning of evolution. In Fig.~\ref{fig:adv_diff2D}, we have plotted the evolution for a 2D advection-diffusion equation with shear and rotation transport fields with periodic boundary conditions for a grid size $N_x=N_y=1024$. The simulation runs for time $T=128$ with $dt = 0.2$ and $R=4$. The total qubits required to implement this evolution are 23, where qubits representing the grid are $n_x=n_y=10$ and the ancilla qubits $n_p=3$. The relative $l_2$-norm error for the 2D shear and rotational transport field are $\mathbf{e}_{0.2}=8.39e^-3$ and $\mathbf{e}_{0.2}=1.30e^-2$ respectively. 

\subsubsection{Three dimensional Shear}
\begin{figure}[t]
    \includegraphics[width=\linewidth]{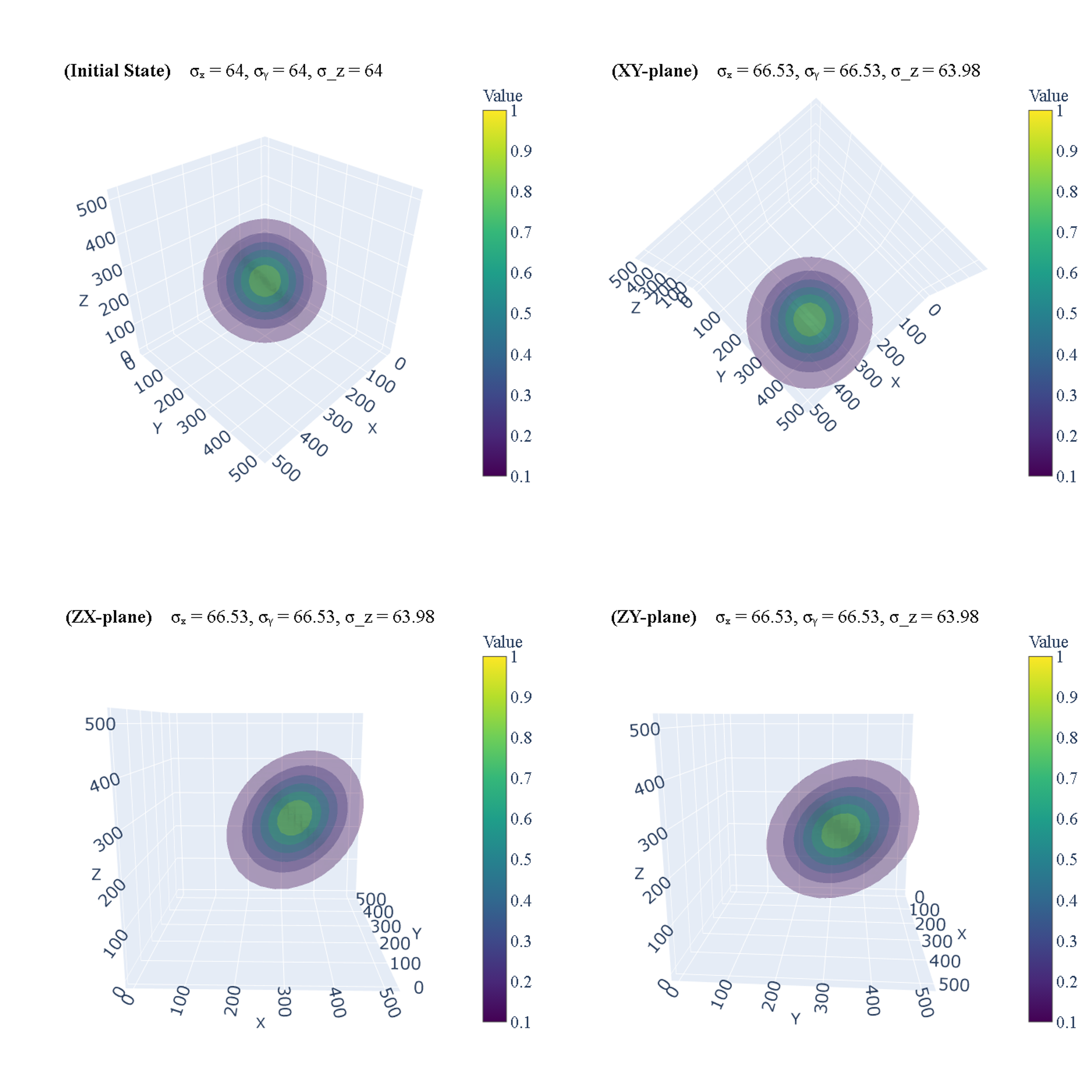}    \caption{\textbf{Simulation results of the advection equation with periodic BC in a $\mathbf{512\times 512 \times 512}$ (27 qubits) lattice}: The initial state is plotted in (a), which evolves according to the 3D advection-diffusion equation with transport field $(\frac{z}{L/2},\frac{z}{L/2},0)$ with diffusion $D=1$. The evolved state from different view angles are plotted in (b), (c), and (d), which are obtained by running the simulation for time $T=64$ with $dt = 0.2$.  The diffusion effect is seen in the enlarge value of the variance of the Gaussian which corresponds to the spread.} 
    \label{fig:3Dadv_diff}
\end{figure}

Lastly, we consider a three-dimensional (3D) advection-diffusion equation with periodic boundary conditions and transport field $v=(\frac{z}{L/2},\frac{z}{L/2},0)$ 
\begin{align}
\begin{cases}\label{Eq:3Dshear_adv_diff}
    \frac{\partial u(x,y,z,t)}{\partial t} + \frac{z}{L/2}\frac{\partial u(x,y,t)}{\partial x}+ \frac{z}{L/2}\frac{\partial u(x,y,t)}{\partial y} = D \nabla^2u(x,y,z,t),  \\
    u(x,y,z,t) = u(x+L,y,z,t),  \\ u(x,y,z,t) = u(x,y+L,z,t),\\ u(x,y,t) = u(x,y,z+Lt),
\end{cases}
\end{align}
where the computation domain is given by $x,y,z\in [0,L]$ and the $D=1$. The 3D diffusion term has the following form 
\begin{align}
&\nabla^2u(x,y,z,t) \nonumber \\ &= D\left(\frac{\partial^2 u(x,y,t)}{\partial x^2} + \frac{\partial^2 u(x,y,t)}{\partial y^2}+\frac{\partial^2 u(x,y,t)}{\partial z^2}\right).    
\end{align}
We take the initial state as a 3D Gaussian 
\begin{align}
    &u_{3Dshear}(x,y,z,0)\nonumber \\&=\frac{1}{\mathcal{N}}\exp\left(-\frac{(x-\mu)^2+(y-\mu)^2+(z-\mu)^2}{2\sigma^2}\right),
    \label{Eq:initial_state_3D}
\end{align}
where $\mu=L/2$ and $\sigma=L/8$. Under evolution with Eq.~\eqref{Eq:3Dshear_adv_diff}, the solution at time $T$ is given by 
\begin{align}
    &u_{3Dshear}(x,y,z,T)\nonumber \\&=\frac{1}{\mathcal{N}}\exp\left(-\frac{(x-\frac{z}{L/2}-\mu)^2+(y-\frac{z}{L/2}-\mu)^2+(z-\mu)^2}{2\sigma_T^2}\right),
    \label{Eq:analytical_3Dshear_diff}
\end{align}
where the $\sigma_T$ is defined from Eq.~\eqref{enlargedsigma}.
The evolution due to this transport field results in advection along $x$ and $y$ axes with linearly varying velocities along $z$ direction. This can be implemented with Eq. ~\eqref{Eq:var_vel_algo}, using two such circuits, one along the $x$ and $y$ direction each with normalized velocities $\frac{y}{L/2}$ and $\frac{x}{L/2}$ respectively. We plot the 3D evolution results in Fig.~\ref{fig:3Dadv_diff}, for $N_x=N_y=N_z=9$, i.e, a grid size of $512\times 512 \times 512$, $n_p=3$, $R=4$, $dt=0.1$, and $T=64$. To check the accuracy of the result of the simulated result with respect to the exact evolution, we calculated the relative $\ell_2$-norm error, which comes out to be $\mathbf{e}_{0.1}=1.12e^{-2}$. 30 total qubits are required to run this simulation of which $27$ qubits are for the physical grid and $3$ ancilla qubits.

\textbf{Note}: There are further improvements possible in the simulation results for all the cases presented above by having more mesh points both in $x$ and $p$, which requires setting up a problem with a bigger system size with $n_x$ and more ancilla qubits $n_p$. We refer the reader to \cite{Hu2024quantumcircuits} for the analysis on how the system and ancilla qubits affect the accuracy of the simulation since it is applicable in our case also.

\begin{figure*}[t]
    \includegraphics[width=\linewidth]{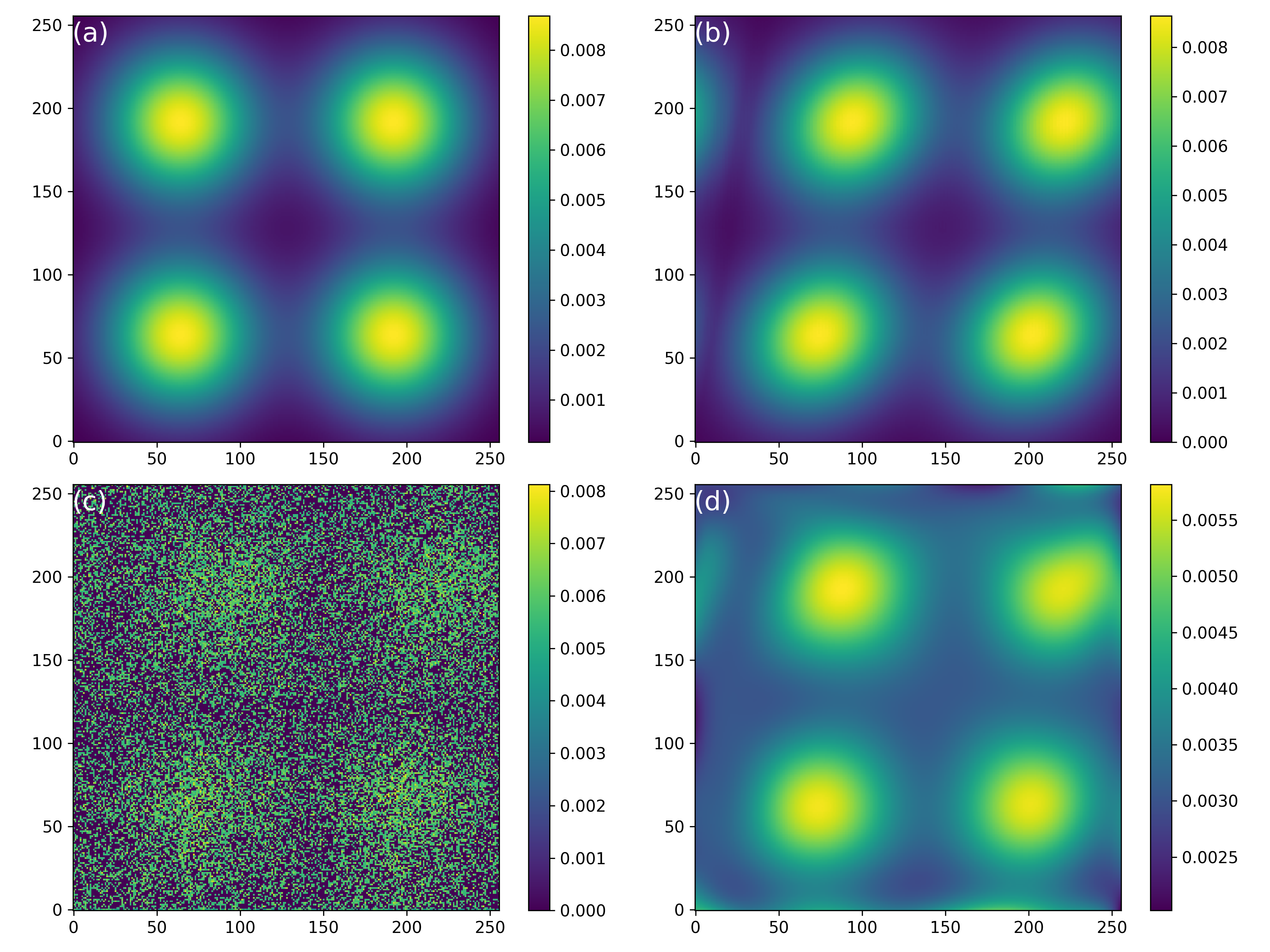} 
    \caption{\textbf{ Hardware results for ($\mathbf{256\times256}$) grid points in two dimensions (16 qubits) : }.   The color contour plot represents the values of the scalar field $u(t)$ moving with a transport $(x,0)$ at each grid point. 
of the scalar field. (a) The initial scalar field at $t=0$. Results for $u(t=20.0)$, dt = 1.0 at statevector level, and hardware results are shown in (b) and (c), respectively. (d) is obtained after applying Savitzky–Golay filter to the noisy data in (c).  It clearly shows that the scalar field could be reconstructed from finite number of shots (100,000).  }
    \label{fig:hardware_res}
\end{figure*}

\section{Hardware results}
In order to demonstrate our algorithm, we run our computations for several time steps.  This amounts to running the circuits for eq 15 in IBM’s 156-qubit processor ibmq\_fez based on the Heron architecture.  
Given the complexity of the upwinding scheme, the circuit depth for our quantum simulation even for one time step could be gigantic. In order to run them on the hardware, substantial transpilation and approximation were implemented without compromising  much of the accuracy. We describe the methods here.  

\subsubsection{Circuit construction and error suppression}

The initial state for the simulation is constructed using Qiskit's \texttt{prepare\_state()}. When a Statevector argument is passed in form of an array, this function prepares the state based on the Isometry synthesis \cite{iten2016quantum}.
Like all Hamiltonian simulation algorithms, our approach also experiences increasing circuit depth with each time step. For a 16-qubit problem ($8\times 8$ lattice), the circuit depth for a single time step reaches approximately $100,000$. To mitigate this issue, we first extract the statevector of the quantum circuit at each time step and reconstruct it using Qiskit's \texttt{initialize()} function. This approach helps keep the circuit depth limited to that of a single time step.

Once we construct the circuit for the full simulation, we apply an additional round of optimization using approximate quantum compilation with tensor networks (AQC-Tensor) to achieve a lower circuit depth than would typically be required for the entire evolution. Specifically, we generate a target tensor-network state for the full circuit using the Qiskit Aer matrix-product state (MPS) simulator, currently the only supported tensor network backend. We then construct a matrix-product representation of the state, which AQC approximates, which we call $\ket{u}_{AQC}$.

Next, we build a general, parameterized ansatz circuit using Qiskit's \texttt{RealAmplitude}, where the optimized parameters yield the tensor network circuit. To achieve this, we minimize the simplest possible cost function, \texttt{MaximizeStateFidelity}, using the L-BFGS optimizer from SciPy. We will call this ansatz state as $\ket{\Tilde{u}(\Theta)}$, where $\Theta := \{ \theta_0, \theta_1, ..\},\; \theta_{j} \in \mathbb{R}$. For the results shown in this work, we use $\abs{\ip{\Tilde{u}(\Theta)}{u_{AQC}}}>0.99$. In other words,  our ansatz has $> 99 \%$  fidelity with the AQC approximate state. This step enables us to get down our circuit depth from $\sim 100,000$ to $500$. 
Starting from these reconstructed circuits of $\ket{u}_{AQC}$, we compile the most efficient version using Qiskit's pass manager with an optimization level of two. Additionally, we apply dynamical decoupling using standard Qiskit tools by incorporating periodic gate sequences, aiming to mitigate undesired system-environment interactions affecting idle qubits \cite{ezzell2023}.
\subsubsection{Error mitigation}
\label{sec:error_mitigation}
 We implemented two error mitigation techniques based solely on the observed characteristics of the hardware output. The first approach stems from noticing that the histogram of noisy bitstrings includes contributions from less relevant states, which are not present in the ideal, noise-free distribution \cite{gomes2023computing}. These extraneous contributions reduce the weight of more meaningful basis states. To address this, we eliminate bitstrings with counts below a threshold $\epsilon_{th}$ by setting their counts to zero and then redistribute those counts evenly among the more significant states. Furthermore, since we are analyzing a shear operation in two dimensions, we anticipate an increase in gate usage, and consequently, noise, as we move farther from the $X=0$ axis. To account for this spatial variation, the threshold is not kept constant but scaled with position as $\epsilon_{th} = y\epsilon_{cut}$, where $\epsilon_{cut} = 1e^{-5}$.


Although the initial step improves the results, the mitigated data still suffers from the fact that the number of measurements (shots) for the full state tomography is finite. 
So, we apply the Savitzky–Golay technique to smooth the fluctuations \cite{savitzky1964smoothing}. Commonly used in digital signal processing, this filter smooths a dataset to improve accuracy without altering the overall shape of the signal. 
In our work, we apply the method in both $x$ and $y$ directions. The final results are discussed in the next section.


\subsection{Results}
We run our hardware experiments for a $256\times 256$ grid lattice moving under a transport field $v = (y/L,0)$. For a lattice of that size, we need only $8+8$ that is 16 qubits.  We have used two ancilla qubits ($n_p = 2$) and for advection and $n_p=3$ for advection-diffusion and $R=8$ in order to incorporate the warped transformation. Following the discussion in section \ref{sec:circuit_approx}, we can ignore the $c-V_1$ term and keep only the inverse of $H_1$ and $H_2$ in order to maintain the upwinding scheme for advection. However, for diffusion we keep all terms. We keep the effect of the number ancilla qubit on the accuracy for future research. 

We assume PBC and an initial field which is a superposition of two gaussian in each direction,
\begin{align}
    u(x,y,0) = \frac{1}{\mathcal{N}}\sum_{i,j =\{1,2\}}\exp\left(-\frac{(x-\mu_i)^2+(y-\mu_j)^2}{2\sigma^2}\right) \nonumber,
\end{align}

where $\mathcal{N}$ is a normalization factor.
We have chosen $\mu_1 = L/2-L/4$, $\mu_2 = L/2+L/4$ and $\sigma = L/8$. We run the simulation upto $T=40$ for $dt=1.0$, that is for 40 Trotter steps.  The results are shown in Fig~\ref{fig:hardware_res} (a-d). Figure ~\ref{fig:hardware_res} (a) shows the initial state $u(x,y,0)$, (b) is the statevector result and (c) is the raw data obtained from a full state tomography from the hardware. Since $u$ is real, the complex phase factor is irrelevant here.  We take a square root and normalize the counts to obtain the coefficients $c_{x,y}$ in eq.~\eqref{eq:statevector} and plot them in (c) after reshaping them as a $(2^{n_x},2^{n_y})$ array.  In (d) we show the hardware results after performing the error mitigation schemes described in section \ref{sec:error_mitigation}. The $\ell_2$ norm error for our hardware result is $\mathbf{e} = 1.96$.

\begin{figure}
    \centering
     \includegraphics[width=\linewidth]{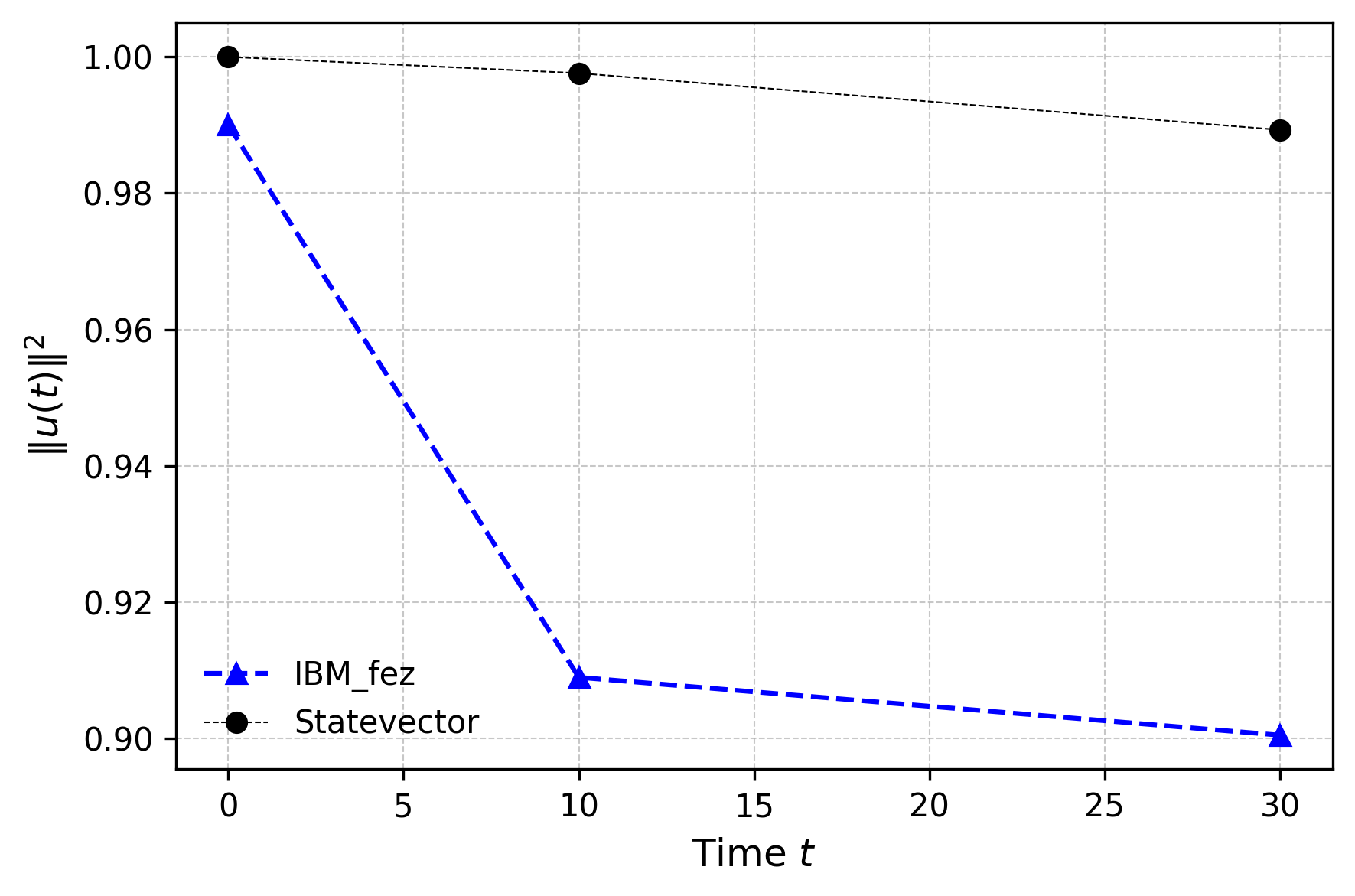} 
    \caption{ Hardware results for advection diffusion for a $256\times 256$ grid in 2D. We have used 19 qubits (16 system qubits and 3 ancilla qubits). The plot shows the evolution of $\norm{\ket{u(t)}}^2$ which is obtained by measuring the observable $\hat{O}$ in eq.~\eqref{eq:projection} } 
    \label{fig:adv-diff-hardware}
\end{figure}

The results for advection-diffusion is shown in Fig~\ref{fig:adv-diff-hardware}. Following \cite{Hu2024quantumcircuits}, we present the total energy $\norm{u}^2$ by measuring the observable $\hat{O}$ for different final time $T\in\{0,10,30\}$. $\hat{O}$ is defined as,
\be 
\hat{O} = I^{\otimes (n_{x}+n_{y})}\otimes\dyad{1}{1}\otimes\dyad{0}{0}^{\otimes (n_p - 1)}
\label{eq:projection}
\ee 
which projects the full quantum state (including the ancilla) $\ket{\hat{\textbf{v}}}(T)$ on to $\ket{u(T)}$. The result is show in fig.~\ref{fig:adv-diff-hardware}.

\section{Conclusion and future work}
 We have developed a new scalable quantum algorithm to include non-zero vorticity into the advection-diffusion equation. In the advection component of the equation, we incorporated non-trivial transport fields using an upwinding discretization scheme and demonstrated it for shear and rotational transport fields on an emulator. The simulation results scale up to 30 qubits grid size for the shear flow in 3D. We have also proposed a few approximations to the upwinding scheme that reduce the depth of the quantum circuit and avoid the use of ancilla qubits and mid-circuit measurements. This makes the algorithm more NISQ-friendly. Finally, we have demonstrated our method on a 16-qubit system which is equivalent to a grid of $256\times256$ lattice on a real quantum hardware for advection and a 19-qubit system to simulate advection and diffusion for the same grid size. Since diffusion is a non-conservative process we have used three ancilla qubits. 
 
 The outcome of a quantum computer simulation is ultimately encoded in a quantum state, as described in Eq.~\eqref{eq:encoding}. In general, extracting full information—such as all quantum amplitudes—incurs exponential cost, making this approach impractical for large systems. Therefore, it is crucial to focus on ‘macroscopic’ observables, such as averages or higher moments of dynamic variables, which can be efficiently estimated from the final state. In future work, we plan to explore observables like flux or kinetic energy \cite{Hu2024quantumcircuits}, whose expectation values can be accessed without such overhead. Additionally, the recently introduced concept of classical shadows \cite{huang2020predicting} offers a promising framework for efficiently retrieving essential features from the quantum output.

As a natural progression of this work, we aim to extend our framework to accommodate more complex transport fields, including time-dependent and turbulent flows. Future efforts will also focus on scaling the algorithm to a larger number of qubits through advanced quantum circuit optimization techniques. Beyond fluid dynamics, we envision applying our methods to a broader class of partial differential equations, such as the heat equation, Maxwell’s equations, and elasticity problems involving stress and strain. Ultimately, our long-term objective is to enable the simulation of general-purpose PDEs through Hamiltonian-based quantum algorithms.
 
 
\subsubsection*{Acknowledgement} The authors thank Chris Hill and Hamed Mohammadbagherpoor for useful discussions.

\appendices 

\section{Circuits for implementing $\tilde{V}_D(\tau)$, $\tilde{V}_1(t)$ and $\tilde{V}_2(t)$}\label{appendix:circuitsv1v2}
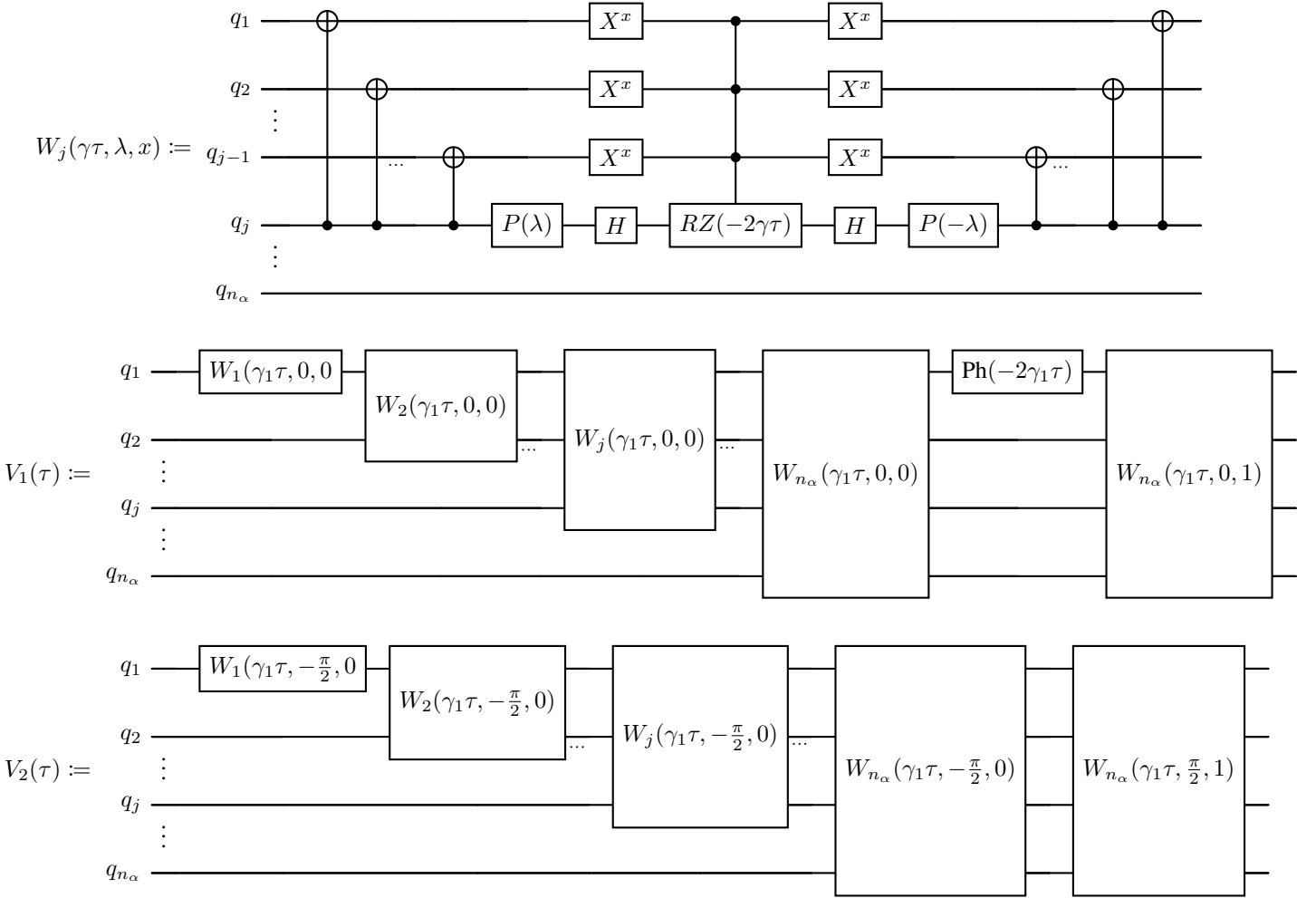
\begin{figure*}[htbp]
    \centering
    \[
    W_j(\gamma\tau, \lambda, x) \coloneqq   
        \begin{quantikz}[row sep={1cm,between origins}, column sep=0.4cm]
        \lstick{$q_1$} & \qw &  \targ{} & \qw & \qw & \qw & \qw & \gate{X^{x}} & \ctrl{3} & \gate{X^{x}} & \qw & \qw & \qw & \qw &\targ{} &\qw \\
        \lstick{$q_2$} & \qw{\vdots}  & \qw & \targ{} & \qw & \qw & \qw & \gate{X^{x}} & \control{} & \gate{X^{x}} & \qw & \qw & \qw & \targ{} &\qw & \qw \\
        \lstick{$q_{j-1}$} & \qw &  \qw & \qw &  \qw{\ldots} & \targ{} & \qw & \gate{X^{x}} & \control{} & \gate{X^{x}} & \qw & \targ{} & \qw{\cdots} & \qw & \qw & \qw \\
        \lstick{$q_{j}$} & \qw{\vdots}  & \ctrl{-3} & \ctrl{-2} & \qw & \ctrl{-1} & \gate{P(\lambda)} & \gate{H} & \gate{RZ(-2\gamma\tau)} & \gate{H} & \gate{P(-\lambda)} & \ctrl{-1} & \qw & \ctrl{-2} & \ctrl{-3} & \qw\\
        \lstick{$q_{n_{\alpha}}$} & \qw & \qw &  \qw & \qw & \qw & \qw & \qw & \qw & \qw& \qw & \qw & \qw & \qw & \qw & \qw
    \end{quantikz}
    \]

    \[ 
    V_1(\tau)\coloneqq    
    \begin{quantikz}[row sep={1cm,between origins}, column sep=0.35cm]
        \lstick{$q_1$} & \qw & \gate{W_1(\gamma_1\tau, 0, 0}&  \gate[2]{W_{2}(\gamma_1\tau, 0, 0)} & \qw & \gate[3]{W_{j}(\gamma_1\tau, 0, 0)} & \qw & \gate[4]{W_{n_{\alpha}}(\gamma_1\tau, 0, 0)} & \gate{\text{Ph}(-2\gamma_{1}\tau)} & \gate[4]{W_{n_{\alpha}}(\gamma_1\tau, 0, 1)} & \qw \\
        \lstick{$q_2$} & \qw{\vdots}  & \qw &  & \qw{\cdots} &  & \qw{\ldots} &  & \qw & & \qw \\
        \lstick{$q_{j}$} & \qw{\vdots} & \qw & \qw & \qw &  & \qw &  & \qw &  & \qw \\
        \lstick{$q_{n_{\alpha}}$}  & \qw & \qw & \qw & \qw & \qw & \qw &  & \qw &  & \qw 
    \end{quantikz}\]
    
     \[ 
     V_2(\tau)\coloneqq     
     \begin{quantikz}[row sep={1cm,between origins}, column sep=0.35cm]
        \lstick{$q_1$} & \qw & \gate{W_1(\gamma_1\tau, -\frac{\pi}{2}, 0} &  \gate[2]{W_{2}(\gamma_1\tau, -\frac{\pi}{2}, 0)} & \qw & \gate[3]{W_{j}(\gamma_1\tau, -\frac{\pi}{2}, 0)} & \qw & \gate[4]{W_{n_{\alpha}}(\gamma_1\tau, -\frac{\pi}{2}, 0)} & \qw & \gate[4]{W_{n_{\alpha}}(\gamma_1\tau, \frac{\pi}{2}, 1)} & \qw \\
        \lstick{$q_2$} & \qw{\vdots}  & \qw &  & \qw{\cdots} &  & \qw{\ldots} &  & \qw & & \qw \\
        \lstick{$q_{j}$} & \qw{\vdots} & \qw & \qw & \qw &  & \qw &  & \qw &  & \qw \\
        \lstick{$q_{n_{\alpha}}$}  & \qw & \qw & \qw & \qw & \qw & \qw &  & \qw &  & \qw 
    \end{quantikz}\]
    
    \caption{Quantum circuits for $W_j(\gamma\tau, \lambda, x), V_1(\tau) \quad \text{and} \quad  V_2(\tau)$.}
    \label{fig:basic_circuits}
\end{figure*}

In this section we present the circuits required for implementing $\tilde{V}_D(\tau)$, $\tilde{V}_1(\tau)$ and $\tilde{V}_2(\tau)$. $\tilde{V}_1(\tau)$ and $\tilde{V}_D(\tau)$ have the same circuit, with the only difference being the input $\gamma_D$ and $\gamma_1$ values. Due to this, we will present the circuits for $\tilde{V}_1(\tau)$ only.  As described in the circuits in Fig.~\ref{fig:basic_circuits}, $V_1(\tau)$ and $V_2(\tau)$ are implemented via the unitary operator $W_j(\gamma\tau, \lambda, x)$ that takes as input $\gamma$ and $\tau=dt$, $\lambda$ and $x\in \{0,1\}$. The full circuit for $\tilde{V}_1(\tau)$ and $\tilde{V}_2(\tau)$ are then implemented as 

\begin{align*}
   \tilde{V}_1(\tau) \coloneqq \prod_{\alpha=1}^d (V_1(\tau))_{\alpha},\\
   \tilde{V}_2(\tau) \coloneqq \prod_{\alpha=1}^d (V_2(\tau))_{\alpha}.
\end{align*}

For more details on how the individual circuits for $W_j(\gamma\tau, \lambda, x)$, $V_1(\tau)$, and $V_2(\tau)$ are obtained, we refer the reader to \cite{Hu2024quantumcircuits} from which these are inspired.

\bibliographystyle{unsrt}
\bibliography{references}

\end{document}